\documentclass[prl,noeprint,twocolumn,floatfix,nobibnotes]{revtex4-1}

\usepackage{amsmath,amssymb,wscmath}
\usepackage{color,framed}
\usepackage[pdftex]{hyperref,graphicx}
\hypersetup{colorlinks = true, urlcolor = blue, linkcolor = blue, citecolor = blue}

\begin{document}
\title{Ising quantum criticality in Majorana nanowires}
\author{William S. Cole}
\email{wcole1@umd.edu}
\author{Jay D. Sau}
\author{S. Das Sarma}
\affiliation{Condensed Matter Theory Center / Joint Quantum Institute / Station Q Maryland \\ University of Maryland, College Park, MD 20742, USA}
\begin{abstract}
Finite-length one-dimensional topological superconductor wires host localized Majorana zero modes at their ends. In realistic models, these appear only after a topological quantum critical point is crossed by external tuning of parameters. Thus, there is a universal finite-size scaling, governed by the critical point, that dictates the evolution of the energy of the Majorana modes near the transition. We first describe this scaling, then apply it in detail to an explicit synthetic topological superconductor model. Our work not only connects Ising quantum criticality with realistic nanowires in the presence of spin-orbit coupling, Zeeman splitting and superconductivity, but also provides a viable experimental route for discerning the existence of the topological quantum critical point.
\end{abstract}
\date{\today}
\maketitle

An intriguing consequence of topological order in matter is the existence of a nearly-degenerate ground state subspace, in which matrix elements of local operators (and therefore energy splittings between ground states) fall off exponentially with growing system size. This ``exponentially protected" degeneracy is a direct manifestation of the nonlocal topological nature of the system, as it does not arise from any obvious symmetry of the Hamiltonian~\cite{tqc_review}. A practical motivation underlying the recent surge of interest in topological order is the prospect of storing quantum information in nonlocal degrees of freedom (the operators that transform different ground states into one another) to make that storage robust against local perturbations~\cite{kitaevAOP}. Topological superconductor (TS) wires, unique in their simplicity and experimental feasibility, support a two-dimensional nearly-degenerate ground state space, containing one ``even" and one ``odd" fermion-number-parity ground state, with the appropriate parity-switching operator shared non-locally between so-called ``Majorana modes" localized at the wire ends~\cite{kitaev_wire, nqi_review}. Recent experiments~\cite{copenhagen_exppro} are believed to provide the first demonstration of the expected exponential-in-length suppression of the splitting between the two parity ground states (i.e. ``exponential protection") at high magnetic fields, after passing through a field-tuned quantum critical point (the topological quantum phase transition, or TQPT~\cite{readgreenTQPT}), which is required to establish TS.
Motivated by such spectroscopic probes of the exponential scaling of presumed Majorana modes, we investigate here a deeper idea: the relationship between the exponential scaling indicative of topological order and the \emph{universal finite-size scaling} (FSS) \cite{fisherbarber} that one expects to observe in real nanoscale devices near a quantum critical point - the aforementioned QCP being, in principle, a generic feature of all realistic 1D Majorana systems. The Majorana TQPT in this problem should have identical properties to the quantum transverse field Ising model~\cite{pfeuty_tfim}, and hence, the Majorana modes are sometimes referred to as Ising anyons (with ``anyons" distinguishing the fact that these modes are neither fermions nor bosons as they obey an exotic non-Abelian braiding statistics~\cite{tqc_review}).

Our starting point is the widely-used single band free fermion model of a one-dimensional, spin-split, spin-orbit coupled nanowire \cite{lutchyn_wire, oreg_wire}, with the single-particle hamiltonian
\be
h(x) = \left(-\frac{\partial^2_x}{2m^*} - \mu \right) + i \alpha \sigma^y \partial_x + V \sigma^x
\ee
in a uniform, local, spin-singlet superconducting pair potential,
\be\elabel{ham}
H = \int_0^L dx \; \left( \psi^\dagger \, h(x) \, \psi + \frac{1}{2}\Delta \psi^{\phd}_{\su} \psi^{\phd}_{\sd} + \mbox{h.c.} \right)
\ee
where the integration limits denote hard-wall confinement in a wire of length $L$. The system defined by \eref{ham} is universally considered to be the appropriate description for realistic Majorana nanowires being extensively studied in many laboratories, and as the physical realization of the Kitaev chain~\cite{kitaev_wire}, albeit with important differences (some of which are discussed below). For any nonzero spin-orbit coupling $\alpha$, this model supports two topologically distinct phases. In the $L \rightarrow \infty$ limit, these are separated by a single critical line $V_c^2 = \Delta^2 + \mu^2$, \fref{pd}a, where the bulk gap closes. This closure of the bulk gap is the defining feature of the existence of a TQPT. The ``topologically nontrivial" phase ($V^2 > \Delta^2 + \mu^2$) is distinguished in part by supporting isolated, localized Majorana modes at the ends of the wire. For finite wires, however, this bulk gap is obstructed from actually vanishing on the erstwhile critical line as a finite-size effect \cite{mishmash_prb}, illustrated in \fref{pd}b. This non-closure of the gap for any real (i.e. finite) system across the putative TQPT is similar to the well-known situation in thermodynamic phase transitions where a correlation length remains non-divergent at a critical point, cut off by the system size (although the thermodynamic phase transition itself is defined by the divergent correlation length). Ref.~\onlinecite{mishmash_prb} discussed the consequences for finite-size spectroscopy of the \emph{bulk} spectrum, expanding around the QCP. Here, we establish that the QCP asserts itself in a universal exponential approach of Majorana modes toward zero energy.

\begin{figure}[t]
\centering
\includegraphics[width=0.48\textwidth]{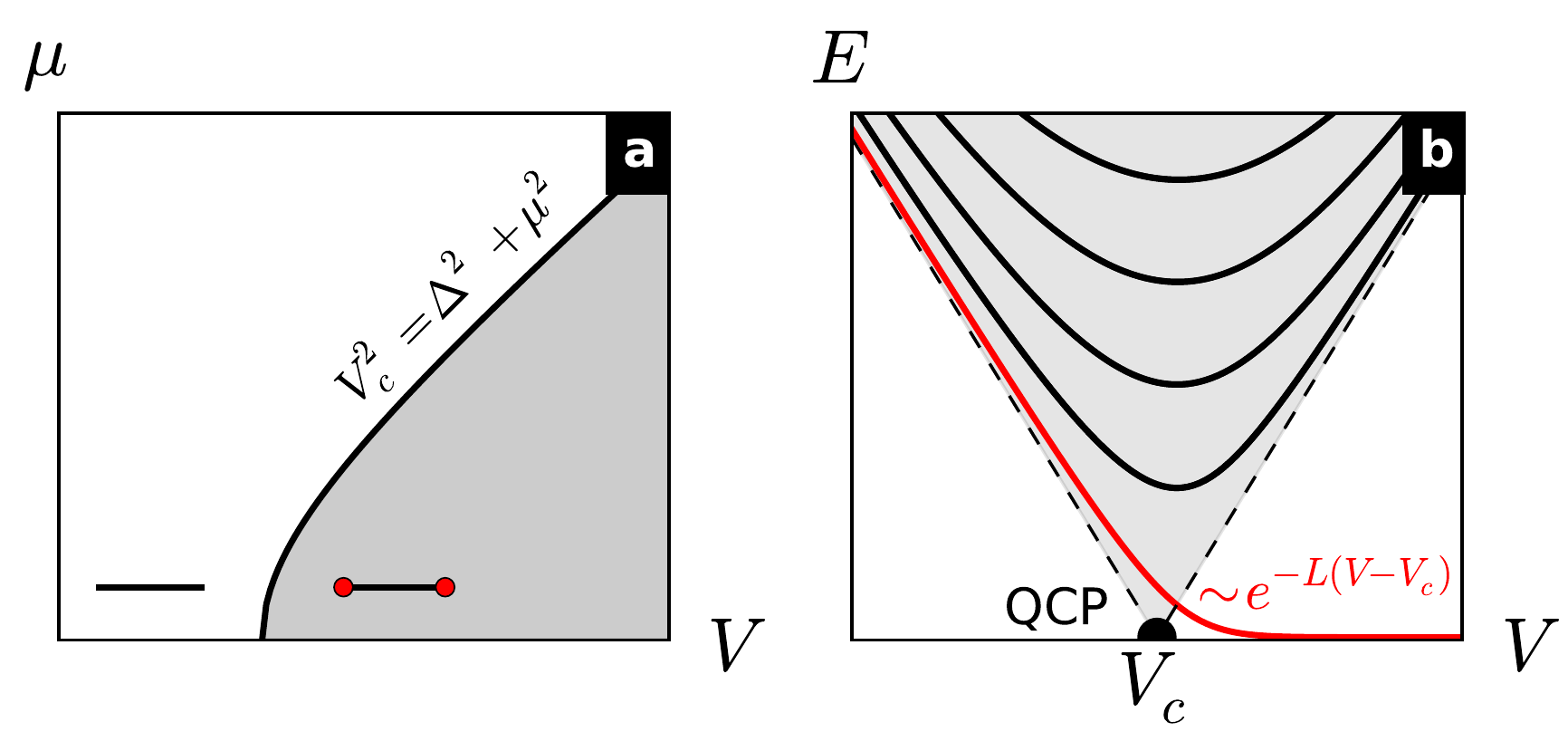}
\caption{
(a) Phase diagram of the model \eref{ham} with a critical line separating topologically trivial and nontrivial parameter spaces. (b) For infinite length wires, the bulk gap closes and reopens linearly at the QCP (dashed lines). For finite length wires, the bulk spectrum (black) is discrete, the gap does not close, and the lowest-lying level (red), representing an inchoate Majorana pair, exponentially approaches zero above the QCP.
}
\flabel{pd}
\end{figure}

\emph{Finite-size scaling from the QCP} ---
In the above-mentioned $L\rightarrow\infty$ gap closing transition, there is a correlation length that goes as $\xi_{\infty} \propto |V-V_c|^{-\nu}$ (focusing on the field-tuned transition obtained by varying $V$, the Zeeman splitting), and correspondingly the lowest-excitation gap goes as $E_{0, \infty} \propto |V-V_c|^{\nu}$. The Kitaev chain~\cite{kitaev_wire}, which is equivalent to the transverse-field Ising model from a criticality perspective, has a universality class characterized by the exponent $\nu=1$~\cite{pfeuty_tfim}. As mentioned, for a finite length wire the excitation gap fails to exactly close. In this work, we will show that the lowest-lying excitation with $E_0 > 0$ (which evolves into a non-local quasiparticle state with support near $x=0,L$) in a finite-size sample approaches zero exponentially with \emph{either} increasing $V$ \emph{or} increasing $L$ for $V > V_c$. This is consistent with a qualitative physical picture where the energy is determined by the overlap of exponentially localized Majoranas at each end with localization length $\propto (V-V_c)^{-\nu}$. The next excited state (which continues to represent the ``bulk gap" in the large $L$ limit) is extended and bounded from below by a confinement gap that scales as $1/L$ even away from $V_c$, as described in Ref.~\onlinecite{mishmash_prb}.

The Ising QCP implies a finite-size scaling hypothesis for the energy of the lowest-lying excitation of a finite-size system~\cite{hamer, cardy_book} (i.e., the ``mass gap" in the Ising theory):
\be\elabel{scaling}
E_0( L ,V ) \simeq L^{-1} \mathcal{F} \left[ L \left( V - V_c \right) \right]
\ee
where $\mathcal{F}$ is a universal scaling function. From this, we can read off that $E_0 \sim 1/L$ exactly at $V = V_c$. However, away from the critical point, the specific dependence of $\mathcal{F}$ on its argument is a characteristic signature of $\nu = 1$ scaling. Our next goal is to understand $\mathcal{F}$ away from the critical point.

To this end, it is known (and recently applied for finite-size scaling of 1D topological phase transitions\cite{mishmash_prb, gulden_prl}) that the appropriate low-energy effective theory near the Ising critical point is that of a single massive Dirac fermion,
\be
\left( -i \sigma^z v \partial_x + \sigma^x m_D(x) - E \right) \psi = 0
\ee
where the velocity $v$ is related to the microscopic parameters of \eref{ham} as described in Ref.~\onlinecite{mishmash_prb}, while the mass is $m_D = m \equiv V - V_c$ inside a length-$L$ well, and $m_D = M$ (with $|M| > |m|$) outside. We require solutions bound to the well with $E < |M|$, but otherwise leave $M$ finite, which could be useful in understanding situations where a finite topological region is embedded in a longer wire (for example, because of disorder~\cite{motrunich_pwave, cole_disorder}), rather than vacuum. A simple calculation yields a condition on $E$ in the form of a transcendental equation
\be\elabel{gencond}
x \coth x + \gamma (L/v)m = 0
\ee
where $x \equiv (L/v)\sqrt{m^2 - E^2}$, which coincides with prior results~\cite{gulden_prl} except for a length-independent finite-$M$ correction
\be
\gamma = (\sgn M) \left( 1 - \frac{E^2}{mM} \right) \left[ 1 - \left( \frac{E}{M} \right)^2 \right]^{-1/2}
\ee

The $E_0$ solution (in fact the entire discrete set of $E$ solutions) to \eref{gencond} can be obtained numerically for any $m$. However, it is useful to identify two limiting cases: at the critical point, $m=0$, the lowest excitation energy is easily obtained in the $M \rightarrow \infty$ limit, $(L/v) E_0 = \pi/2$. (This fails for finite $M$,  however, as $\lim_{m \rightarrow 0}\gamma m \neq 0$.)
We choose the topological case to correspond to positive $m$, negative $M$. For sufficiently large $Lm$, then, there exists an $E_0 \ll m$ solution, such that $E_0$ can be approximated (using $\coth x \simeq 1 + 2e^{-2x}$), leading to an analytic approximation to the desired scaling function:
\be
\mathcal{F} \simeq 2 e^{-(L/v)m} \left( \frac{1}{(L/v)m} - \frac{1}{(L/v)M} \right)^{-1}
\ee
which has the form of \eref{scaling} up to a boundary condition (i.e., $M$) dependent term~\footnote{The $M$ dependence can be made arbitrarily small at fixed $mL$ by increasing $L$, in accordance with our expectation for a non-universal boundary condition contribution to the excitation energy.} and is dominated by the exponential behavior, thus justifying the qualitative overlapping-Majorana interpretation for finite-length wires.

In what follows, we describe detailed numerical tests of the scaling relation in \eref{scaling} explicitly for the lowest eigenvalue of \eref{ham}. Though it comes as no surprise that the critical point of \eref{ham} is in the Ising universality class, \eref{ham} is: (i) nonetheless qualitatively distinct from the Kitaev chain, and not connected to the Ising model by a simple operator transformation, and (ii) a widely used effective continuum model describing real experiments. In this way, it serves as a bridge to experiments on quantum criticality in finite systems; our aim is a simple proof-of-principle simulation of a plausible method to demonstrate Ising universality by analyzing finite-size edge tunneling spectroscopy experiments \cite{copenhagen_exppro, gul_hardgap, deng_qd, frolov_phasediagram}. This requires both the demonstration that a reasonable microscopic model obeys universal scaling, as well as characterization of any non-universal finite size effects. We believe that our finite-size scaling results demonstrate that a careful analysis of the experimental data around the critical magnetic field may be able to compellingly establish the existence of the TQPT.  This should be distinguished from the ``empirical" critical magnetic field, above which a zero bias conductance peak arises in the Majorana nanowire experiments~\cite{copenhagen_exppro, gul_hardgap, deng_qd, frolov_phasediagram}.

\begin{figure}[t]
\centering
\includegraphics[width=0.48\textwidth]{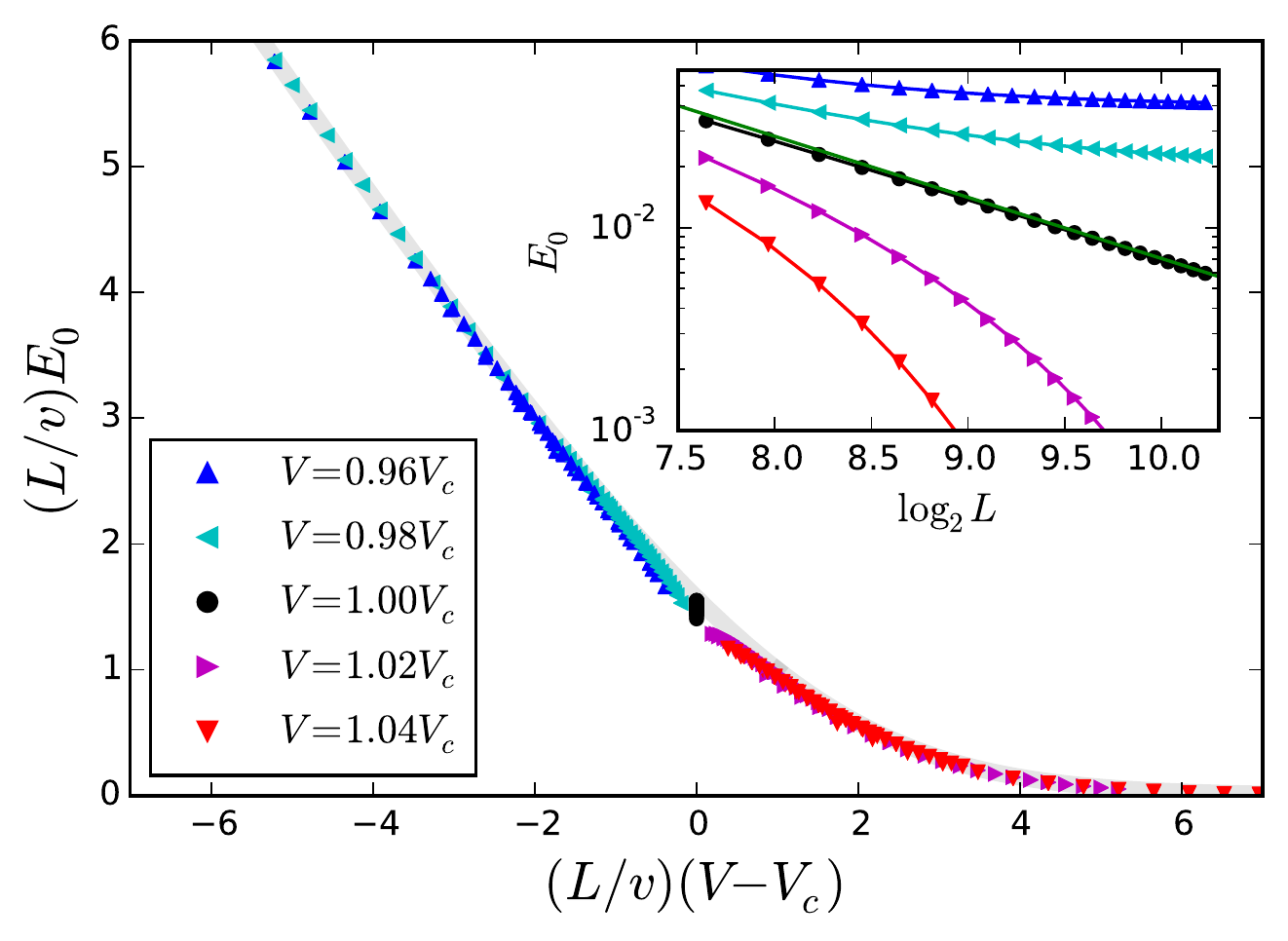}
\caption{
Data collapse on the scaling function \eref{scaling} for $V$ near $V_c$ with $\mu=0$. Exactly at $V_c$, $(L/v)E_0 \sim$ const. for all $L$. Inset: log-log plot of unscaled $E_0(L, V)$, also showing power-law dependence at $V_c$ and non-power-law approach to the large-$L$ limit away from $V_c$.
}
\flabel{scaling_nomu}
\end{figure}

\emph{Case I: Zero chemical potential} ---
First, we restrict to the $\mu = 0$ case.  This is the theoretically ideal situation, although the experimental chemical potential is typically unknown (but the situation is improved by efforts at mapping out the phase diagram of a single device \cite{frolov_phasediagram}). We numerically obtain the smallest eigenvalue of \eref{ham} over a wide range of physical length ($L \sim 0.6-3.6 \mu$m) and spin-orbit parameters ($m^* \alpha^2 \sim 0.1-2 \Delta$). All of the relevant results are contained in the data collapse shown in \fref{scaling_nomu}. We emphasize (i) the collapse on to the universal function $\mathcal{F}$, obtained from \eref{gencond} with $M \rightarrow -\infty$; (ii) the power-law (specifically, $1/L$) dependence of $E_0(L, V_c)$, and non-power-law behavior away from $V_c$; and (iii) the excellent agreement with exponential scaling above the critical point. In the effective theory, the only effect of changing the spin-orbit parameter is a rescaling of the physical length as $L/v$.

Following Ref.~\onlinecite{mishmash_prb}, $v$ is obtained by diagonalizing \eref{ham} in the infinite $L$ limit, expanding the resulting (squared) eigenvalues in $k$ around zero, as is appropriate in the whole critical regime, and then matching to the form
\be\elabel{E2vsK2}
E^2 = m^2 + (vk)^2 + \mathcal{O}(k^4)
\ee
For $\mu = 0$, $v = \alpha$ exactly, while for nonzero $\mu$ (the generic situation) the expression for the Dirac velocity becomes considerably more complicated, with a weak but explicit dependence on $V$. Also, for extremely small systems, one expects additional nonuniversal corrections to $E_0$, in particular $\mathcal{O}(1/L^2)$ contributions absent in the Dirac model. We will show next that even this correction is accidentally small near $\mu = 0$. The combination of these two features contribute to the excellent scaling in \fref{scaling_nomu} even for small $L$.

\emph{Case II: Nonzero chemical potential} ---
We first consider the $\mathcal{O}(1/L^{2})$ corrections to the energy at the critical point,
\be\elabel{L2corr}
E_0 \simeq \frac{\pi v}{2 L} + f(\mu,\alpha)\left( \frac{v}{L} \right)^2
\ee
In \fref{crit_mu} we show the approach to scaling, plotting the numerically obtained $(L/v)E_0$ at the critical point as a function of $(L/v)^{-1}$ with fits to \eref{L2corr} for several different values of $\mu$, as well as plotting $f(\mu)$ for $m^*\alpha^2 = \Delta$. Of particular interest is that $f(\mu, \alpha)$ goes through zero around $\mu \simeq -m^* \alpha^2$, but remains relatively small in the entire range $|\mu| < m^* \alpha^2$. The energy correction can be positive or negative, sensitive to the sign of $\mu$, and can be quite large for realistic $L$.
Away from $V_c$, \eref{scaling} is dominated by the exponential behavior, so $1/L^2$ and higher order corrections should already be strongly suppressed. On the other hand, it is only away from $V=V_c$, and only for nonzero $\mu$, that the effective Dirac velocity $v$ has an explicit dependence on the spin splitting $V$. One might expect this dependence to ruin the scaling relationship \eref{scaling} which is predicated on the effective length of the wire being \emph{independent} of the splitting. This is indeed the case for very weak spin-orbit coupling (i.e., $m^{\ast}\alpha^2 \ll \mu$) or very short wires, because the variation in $V$ required to map out a large range of $(L/v)(V-V_c)$ also becomes large. On the other hand, for even moderate spin-orbit coupling strength we nevertheless find a significant critical region in good agreement with \eref{scaling} (i.e., indicative of $\nu = 1$) even for realistically short $L$, since the dependence of $v$ on $V$ is subleading, resulting in only a very small correction to scaling shown in \fref{off_crit_mu}.

\begin{figure}[t]
\centering
\includegraphics[width=0.48\textwidth]{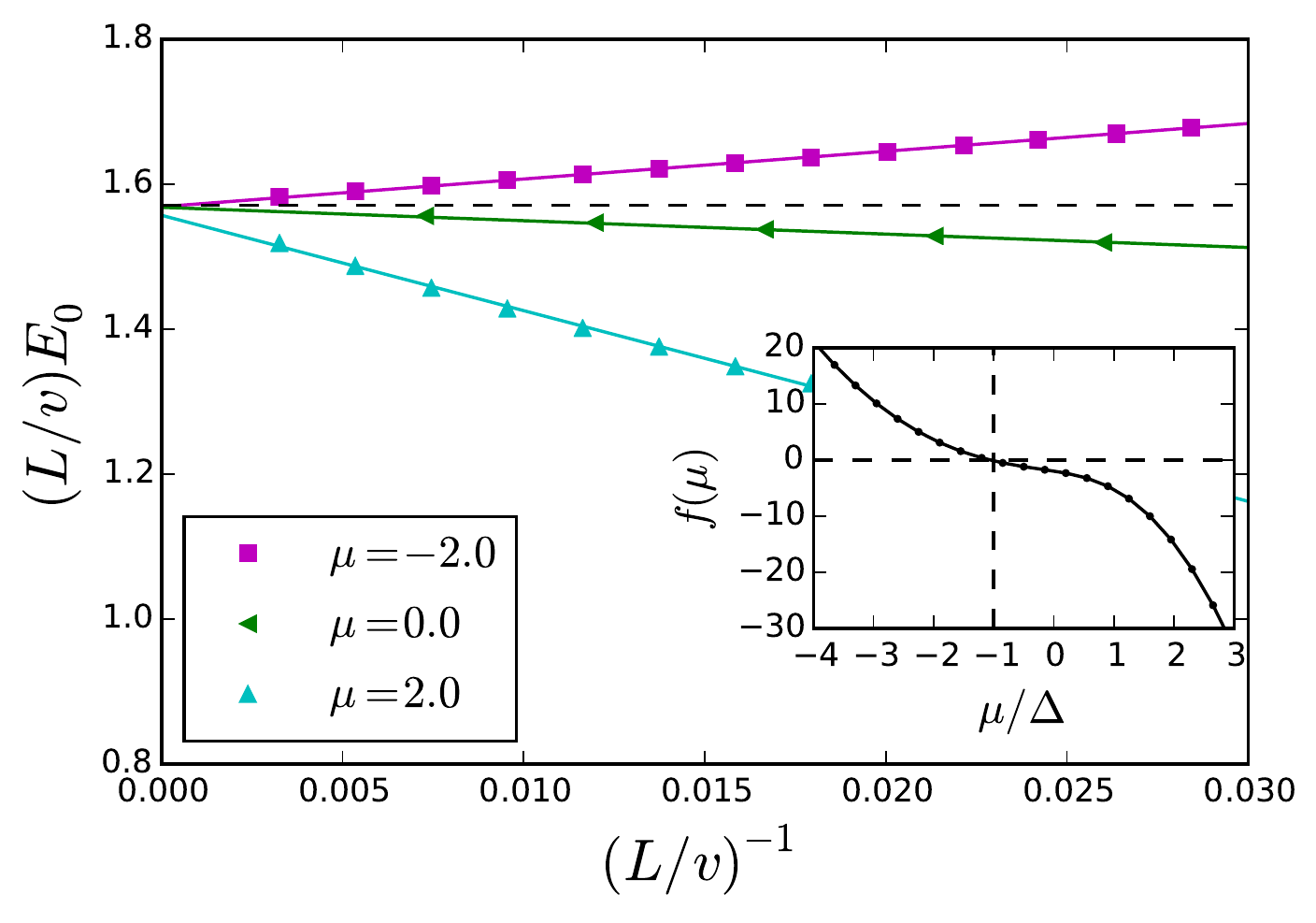}
\caption{
We show the critical ($V=V_c$) gap scaling $(L/v)E_0 \sim \frac{\pi}{2} + \ldots$ for several values of $\mu$, with $m^*\alpha^2 \simeq \Delta$ (relatively strong SOC). All curves converge to the scaling behavior $E_0 = \frac{\pi v}{2 L}$ at large $L$. For shorter $L$, there are systematic $\mu$-dependent corrections. We find good fits to \eref{L2corr}, indicated by solid lines. The slope as a function of $\mu$ is plotted in the inset, and the detailed behavior is described in the text.
}
\flabel{crit_mu}
\end{figure}

\begin{figure}[t]
\centering
\includegraphics[width=0.48\textwidth]{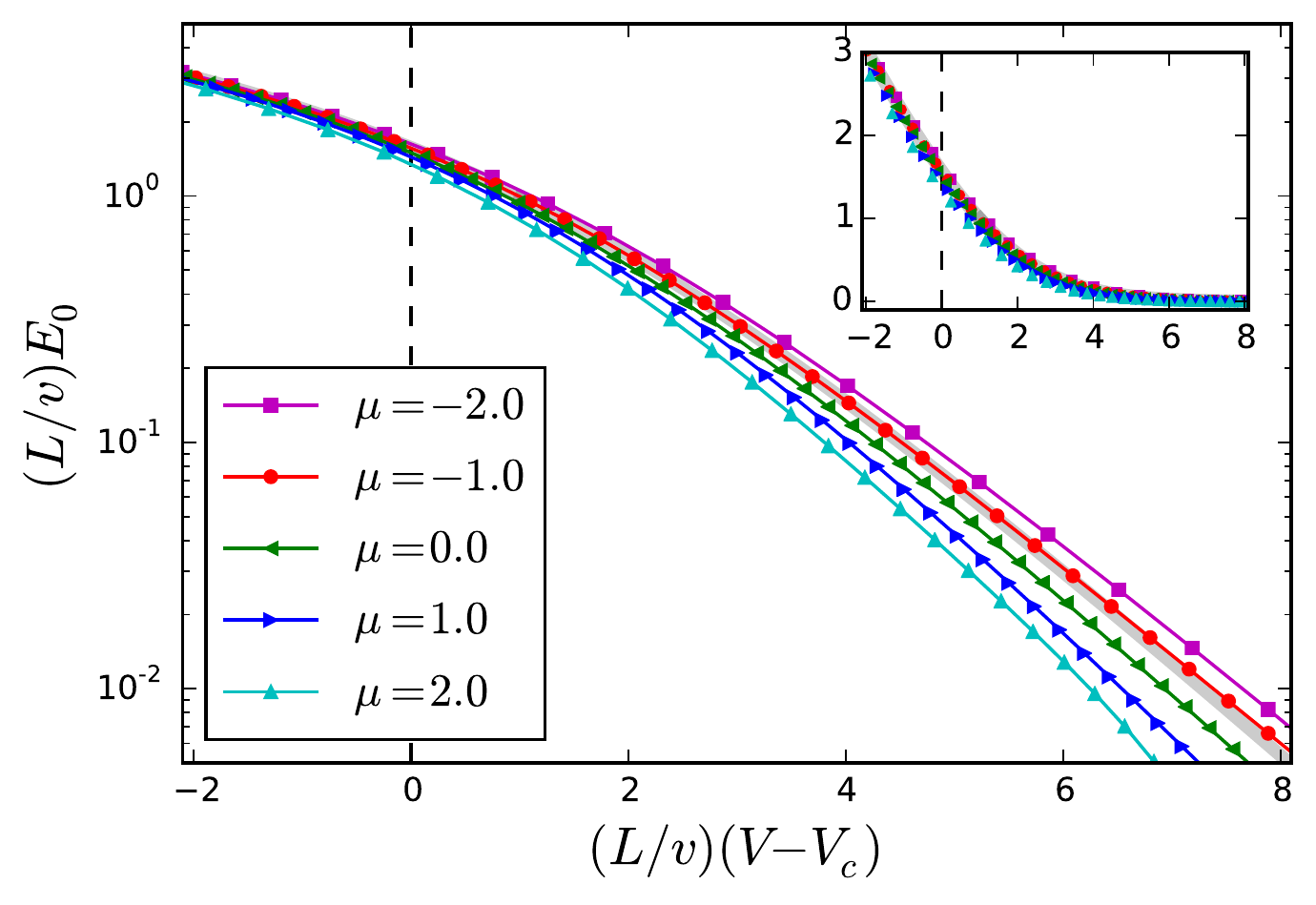}
\caption{
The effect of nonzero chemical potential on scaling far from the critical point, for a fixed length $L = 1$ $\mu$m, $m^*\alpha^2 \simeq \Delta$. Reasonably good data collapse on $\mathcal{F}$ is apparent (see inset). In the exponential regime $V > V_c$, corrections are visible only on a semilog scale; this correction to scaling can be attributed to the weak dependence of the effective Dirac velocity on $V$, which occurs only for nonzero $\mu$.
}
\flabel{off_crit_mu}
\end{figure}

\emph{Discussion} ---
First, we comment briefly on the separate breakdown of scaling behavior that occurs as $V$ increases further past the QCP. The exponential-in-$V$ behavior is known to give way to oscillations with increasing amplitude as $V$ increases. This behavior was studied in detail previously \cite{smokinggun} (and persists qualitatively, if not quantitatively, as other extensions to \eref{ham} are introduced such as weak disorder \cite{diego_realistic}, Majorana overlap mediated by the parent superconductor \cite{zyuzin}, etc.). Oscillations, parity crossings, and other phenomena at high fields are non-universal features of the microscopic model (i.e., they do not occur in the Ising model), however, and are not controlled by the Ising QCP.

The dependence of finite-size scaling here on the chemical potential is subtle and nontrivial as a comparison of \fref{scaling_nomu} to \fref{off_crit_mu} shows, where tuning the chemical potential has a quantitative effect on the scaling function itself (for fixed, relatively short $L$) except very close to the QCP $V=V_c$.  The chemical potential is not quite a dangerous irrelevant variable in the strict sense, but its apparent effect may seem like that of a dangerous irrelevant variable since it suppresses the scaling regime. In this case, one must go closer to the QCP in order to ascertain the correct Ising criticality exactly.

This work establishes universal finite-size scaling of the lowest-lying excitation of a realistic Majorana nanowire model. This provides an independent direct check that the phase transition in this model belongs to the same universality class as the Kitaev chain (i.e., Ising) that was expected based on the shared spectrum of low energy excitations~\cite{lutchyn_wire}.
We find that the role of chemical potential is however surprisingly nontrivial in the finite-size scaling analyses, and nonzero chemical potential introduces severe finite-size corrections that may restrict the scaling regime considerably although the universal Ising criticality is eventually restored approaching the QCP.
More importantly, though, we suggest that detailed spectroscopy through the topological QCP in small systems would provide \emph{experimental} validation of a Majorana interpretation by demonstrating that finite nanowires have the correct scaling properties for this universality class. In addition to the demonstration of universal scaling, we also determined that nonzero chemical potential gives rise to potentially large non-universal contributions to the energy scaling, which could be inverted as a way to estimate the unknown value of $\mu$.

Exponential-in-length scaling as a characteristic signature of Majoranas occurs not just (non-universally) at high fields as in Refs.~\onlinecite{smokinggun, copenhagen_exppro}, but also in the low-field ``foot" of the gap closing near the QCP, which has several practical advantages. (i) Such a foot appears already visible to the eye in several reported spectroscopy experiments, and therefore might be resolvable even in experiments that cannot see any Majorana splitting at high fields. (ii) In such experiments, the exponential behavior is field-tuned - the effective length of the wire is dictated by the distance from the critical point, which is much preferable to comparing samples at different physical but unknown effective lengths. (iii) Presently high fields introduce additional known complications, such as the ubiquitous ``soft gap" above the QCP. This work does not attempt to resolve those problems, however, we note that several groups have demonstrated devices that maintain a hard superconducting gap all the way up to the QCP~\cite{gul_hardgap, deng_qd}, and so relatively low-field evidence for the existence of Majorana modes seems feasible by carrying out finite-size scaling analysis of the existing data in the ``foot" region around the TQPT.  It is germane here to emphasize that so far no experiment has been able to directly detect the QCP (an essentially bulk property) in Majorana nanowires, in spite of several suggestions~\cite{akhmerov_QCP, fregoso_QCP, tewari_QCP}, and so finite-size scaling may turn out to be optimal in establishing the QCP.

This work was supported by Microsoft and LPS-MPO-CMTC.


\begin{thebibliography}{26}%
\makeatletter
\providecommand \@ifxundefined [1]{%
 \@ifx{#1\undefined}
}%
\providecommand \@ifnum [1]{%
 \ifnum #1\expandafter \@firstoftwo
 \else \expandafter \@secondoftwo
 \fi
}%
\providecommand \@ifx [1]{%
 \ifx #1\expandafter \@firstoftwo
 \else \expandafter \@secondoftwo
 \fi
}%
\providecommand \natexlab [1]{#1}%
\providecommand \enquote  [1]{``#1''}%
\providecommand \bibnamefont  [1]{#1}%
\providecommand \bibfnamefont [1]{#1}%
\providecommand \citenamefont [1]{#1}%
\providecommand \href@noop [0]{\@secondoftwo}%
\providecommand \href [0]{\begingroup \@sanitize@url \@href}%
\providecommand \@href[1]{\@@startlink{#1}\@@href}%
\providecommand \@@href[1]{\endgroup#1\@@endlink}%
\providecommand \@sanitize@url [0]{\catcode `\\12\catcode `\$12\catcode
  `\&12\catcode `\#12\catcode `\^12\catcode `\_12\catcode `\%12\relax}%
\providecommand \@@startlink[1]{}%
\providecommand \@@endlink[0]{}%
\providecommand \url  [0]{\begingroup\@sanitize@url \@url }%
\providecommand \@url [1]{\endgroup\@href {#1}{\urlprefix }}%
\providecommand \urlprefix  [0]{URL }%
\providecommand \Eprint [0]{\href }%
\providecommand \doibase [0]{http://dx.doi.org/}%
\providecommand \selectlanguage [0]{\@gobble}%
\providecommand \bibinfo  [0]{\@secondoftwo}%
\providecommand \bibfield  [0]{\@secondoftwo}%
\providecommand \translation [1]{[#1]}%
\providecommand \BibitemOpen [0]{}%
\providecommand \bibitemStop [0]{}%
\providecommand \bibitemNoStop [0]{.\EOS\space}%
\providecommand \EOS [0]{\spacefactor3000\relax}%
\providecommand \BibitemShut  [1]{\csname bibitem#1\endcsname}%
\let\auto@bib@innerbib\@empty
\bibitem [{\citenamefont {{Nayak}}\ \emph {et~al.}(2008)\citenamefont
  {{Nayak}}, \citenamefont {{Simon}}, \citenamefont {{Stern}}, \citenamefont
  {{Freedman}},\ and\ \citenamefont {{Das Sarma}}}]{tqc_review}%
  \BibitemOpen
  \bibfield  {author} {\bibinfo {author} {\bibfnamefont {C.}~\bibnamefont
  {{Nayak}}}, \bibinfo {author} {\bibfnamefont {S.~H.}\ \bibnamefont
  {{Simon}}}, \bibinfo {author} {\bibfnamefont {A.}~\bibnamefont {{Stern}}},
  \bibinfo {author} {\bibfnamefont {M.}~\bibnamefont {{Freedman}}}, \ and\
  \bibinfo {author} {\bibfnamefont {S.}~\bibnamefont {{Das Sarma}}},\ }\href
  {\doibase 10.1103/RevModPhys.80.1083} {\bibfield  {journal} {\bibinfo
  {journal} {Reviews of Modern Physics}\ }\textbf {\bibinfo {volume} {80}},\
  \bibinfo {pages} {1083} (\bibinfo {year} {2008})}\BibitemShut {NoStop}%
\bibitem [{\citenamefont {{Kitaev}}(2003)}]{kitaevAOP}%
  \BibitemOpen
  \bibfield  {author} {\bibinfo {author} {\bibfnamefont {A.~Y.}\ \bibnamefont
  {{Kitaev}}},\ }\href {\doibase 10.1016/S0003-4916(02)00018-0} {\bibfield
  {journal} {\bibinfo  {journal} {Annals of Physics}\ }\textbf {\bibinfo
  {volume} {303}},\ \bibinfo {pages} {2} (\bibinfo {year} {2003})}\BibitemShut
  {NoStop}%
\bibitem [{\citenamefont {{Kitaev}}(2001)}]{kitaev_wire}%
  \BibitemOpen
  \bibfield  {author} {\bibinfo {author} {\bibfnamefont {A.~Y.}\ \bibnamefont
  {{Kitaev}}},\ }\href {\doibase 10.1070/1063-7869/44/10S/S29} {\bibfield
  {journal} {\bibinfo  {journal} {Physics Uspekhi}\ }\textbf {\bibinfo {volume}
  {44}},\ \bibinfo {pages} {131} (\bibinfo {year} {2001})}\BibitemShut
  {NoStop}%
\bibitem [{\citenamefont {{Das Sarma}}\ \emph {et~al.}(2015)\citenamefont {{Das
  Sarma}}, \citenamefont {Freedman},\ and\ \citenamefont {Nayak}}]{nqi_review}%
  \BibitemOpen
  \bibfield  {author} {\bibinfo {author} {\bibfnamefont {S.}~\bibnamefont {{Das
  Sarma}}}, \bibinfo {author} {\bibfnamefont {M.}~\bibnamefont {Freedman}}, \
  and\ \bibinfo {author} {\bibfnamefont {C.}~\bibnamefont {Nayak}},\ }\href
  {\doibase 10.1038/npjqi.2015.1} {\bibfield  {journal} {\bibinfo  {journal}
  {npj Quantum Information}\ }\textbf {\bibinfo {volume} {1}},\ \bibinfo
  {pages} {15001} (\bibinfo {year} {2015})}\BibitemShut {NoStop}%
\bibitem [{\citenamefont {{Albrecht}}\ \emph {et~al.}(2016)\citenamefont
  {{Albrecht}}, \citenamefont {{Higginbotham}}, \citenamefont {{Madsen}},
  \citenamefont {{Kuemmeth}}, \citenamefont {{Jespersen}}, \citenamefont
  {{Nyg{\aa}rd}}, \citenamefont {{Krogstrup}},\ and\ \citenamefont
  {{Marcus}}}]{copenhagen_exppro}%
  \BibitemOpen
  \bibfield  {author} {\bibinfo {author} {\bibfnamefont {S.~M.}\ \bibnamefont
  {{Albrecht}}}, \bibinfo {author} {\bibfnamefont {A.~P.}\ \bibnamefont
  {{Higginbotham}}}, \bibinfo {author} {\bibfnamefont {M.}~\bibnamefont
  {{Madsen}}}, \bibinfo {author} {\bibfnamefont {F.}~\bibnamefont
  {{Kuemmeth}}}, \bibinfo {author} {\bibfnamefont {T.~S.}\ \bibnamefont
  {{Jespersen}}}, \bibinfo {author} {\bibfnamefont {J.}~\bibnamefont
  {{Nyg{\aa}rd}}}, \bibinfo {author} {\bibfnamefont {P.}~\bibnamefont
  {{Krogstrup}}}, \ and\ \bibinfo {author} {\bibfnamefont {C.~M.}\ \bibnamefont
  {{Marcus}}},\ }\href {\doibase 10.1038/nature17162} {\bibfield  {journal}
  {\bibinfo  {journal} {\nat}\ }\textbf {\bibinfo {volume} {531}},\ \bibinfo
  {pages} {206} (\bibinfo {year} {2016})}\BibitemShut {NoStop}%
\bibitem [{\citenamefont {Read}\ and\ \citenamefont
  {Green}(2000)}]{readgreenTQPT}%
  \BibitemOpen
  \bibfield  {author} {\bibinfo {author} {\bibfnamefont {N.}~\bibnamefont
  {Read}}\ and\ \bibinfo {author} {\bibfnamefont {D.}~\bibnamefont {Green}},\
  }\href {\doibase 10.1103/PhysRevB.61.10267} {\bibfield  {journal} {\bibinfo
  {journal} {Phys. Rev. B}\ }\textbf {\bibinfo {volume} {61}},\ \bibinfo
  {pages} {10267} (\bibinfo {year} {2000})}\BibitemShut {NoStop}%
\bibitem [{\citenamefont {{Fisher}}\ and\ \citenamefont
  {{Barber}}(1972)}]{fisherbarber}%
  \BibitemOpen
  \bibfield  {author} {\bibinfo {author} {\bibfnamefont {M.~E.}\ \bibnamefont
  {{Fisher}}}\ and\ \bibinfo {author} {\bibfnamefont {M.~N.}\ \bibnamefont
  {{Barber}}},\ }\href {\doibase 10.1103/PhysRevLett.28.1516} {\bibfield
  {journal} {\bibinfo  {journal} {Physical Review Letters}\ }\textbf {\bibinfo
  {volume} {28}},\ \bibinfo {pages} {1516} (\bibinfo {year}
  {1972})}\BibitemShut {NoStop}%
\bibitem [{\citenamefont {{Pfeuty}}(1970)}]{pfeuty_tfim}%
  \BibitemOpen
  \bibfield  {author} {\bibinfo {author} {\bibfnamefont {P.}~\bibnamefont
  {{Pfeuty}}},\ }\href {\doibase 10.1016/0003-4916(70)90270-8} {\bibfield
  {journal} {\bibinfo  {journal} {Annals of Physics}\ }\textbf {\bibinfo
  {volume} {57}},\ \bibinfo {pages} {79} (\bibinfo {year} {1970})}\BibitemShut
  {NoStop}%
\bibitem [{\citenamefont {{Lutchyn}}\ \emph {et~al.}(2010)\citenamefont
  {{Lutchyn}}, \citenamefont {{Sau}},\ and\ \citenamefont {{Das
  Sarma}}}]{lutchyn_wire}%
  \BibitemOpen
  \bibfield  {author} {\bibinfo {author} {\bibfnamefont {R.~M.}\ \bibnamefont
  {{Lutchyn}}}, \bibinfo {author} {\bibfnamefont {J.~D.}\ \bibnamefont
  {{Sau}}}, \ and\ \bibinfo {author} {\bibfnamefont {S.}~\bibnamefont {{Das
  Sarma}}},\ }\href {\doibase 10.1103/PhysRevLett.105.077001} {\bibfield
  {journal} {\bibinfo  {journal} {Physical Review Letters}\ }\textbf {\bibinfo
  {volume} {105}},\ \bibinfo {eid} {077001} (\bibinfo {year}
  {2010})}\BibitemShut {NoStop}%
\bibitem [{\citenamefont {{Oreg}}\ \emph {et~al.}(2010)\citenamefont {{Oreg}},
  \citenamefont {{Refael}},\ and\ \citenamefont {{von Oppen}}}]{oreg_wire}%
  \BibitemOpen
  \bibfield  {author} {\bibinfo {author} {\bibfnamefont {Y.}~\bibnamefont
  {{Oreg}}}, \bibinfo {author} {\bibfnamefont {G.}~\bibnamefont {{Refael}}}, \
  and\ \bibinfo {author} {\bibfnamefont {F.}~\bibnamefont {{von Oppen}}},\
  }\href {\doibase 10.1103/PhysRevLett.105.177002} {\bibfield  {journal}
  {\bibinfo  {journal} {Physical Review Letters}\ }\textbf {\bibinfo {volume}
  {105}},\ \bibinfo {eid} {177002} (\bibinfo {year} {2010})}\BibitemShut
  {NoStop}%
\bibitem [{\citenamefont {{Mishmash}}\ \emph {et~al.}(2016)\citenamefont
  {{Mishmash}}, \citenamefont {{Aasen}}, \citenamefont {{Higginbotham}},\ and\
  \citenamefont {{Alicea}}}]{mishmash_prb}%
  \BibitemOpen
  \bibfield  {author} {\bibinfo {author} {\bibfnamefont {R.~V.}\ \bibnamefont
  {{Mishmash}}}, \bibinfo {author} {\bibfnamefont {D.}~\bibnamefont {{Aasen}}},
  \bibinfo {author} {\bibfnamefont {A.~P.}\ \bibnamefont {{Higginbotham}}}, \
  and\ \bibinfo {author} {\bibfnamefont {J.}~\bibnamefont {{Alicea}}},\ }\href
  {\doibase 10.1103/PhysRevB.93.245404} {\bibfield  {journal} {\bibinfo
  {journal} {\prb}\ }\textbf {\bibinfo {volume} {93}},\ \bibinfo {eid} {245404}
  (\bibinfo {year} {2016})}\BibitemShut {NoStop}%
\bibitem [{\citenamefont {{Hamer}}\ and\ \citenamefont
  {{Barber}}(1981)}]{hamer}%
  \BibitemOpen
  \bibfield  {author} {\bibinfo {author} {\bibfnamefont {C.~J.}\ \bibnamefont
  {{Hamer}}}\ and\ \bibinfo {author} {\bibfnamefont {M.~N.}\ \bibnamefont
  {{Barber}}},\ }\href {\doibase 10.1088/0305-4470/14/1/024} {\bibfield
  {journal} {\bibinfo  {journal} {Journal of Physics A Mathematical General}\
  }\textbf {\bibinfo {volume} {14}},\ \bibinfo {pages} {241} (\bibinfo {year}
  {1981})}\BibitemShut {NoStop}%
\bibitem [{\citenamefont {Cardy}(1996)}]{cardy_book}%
  \BibitemOpen
  \bibfield  {author} {\bibinfo {author} {\bibfnamefont {J.}~\bibnamefont
  {Cardy}},\ }\href@noop {} {\emph {\bibinfo {title} {Scaling and
  Renormalization in Statistical Physics}}}\ (\bibinfo  {publisher} {Cambridge
  University Press},\ \bibinfo {year} {1996})\BibitemShut {NoStop}%
\bibitem [{\citenamefont {{Gulden}}\ \emph {et~al.}(2016)\citenamefont
  {{Gulden}}, \citenamefont {{Janas}}, \citenamefont {{Wang}},\ and\
  \citenamefont {{Kamenev}}}]{gulden_prl}%
  \BibitemOpen
  \bibfield  {author} {\bibinfo {author} {\bibfnamefont {T.}~\bibnamefont
  {{Gulden}}}, \bibinfo {author} {\bibfnamefont {M.}~\bibnamefont {{Janas}}},
  \bibinfo {author} {\bibfnamefont {Y.}~\bibnamefont {{Wang}}}, \ and\ \bibinfo
  {author} {\bibfnamefont {A.}~\bibnamefont {{Kamenev}}},\ }\href {\doibase
  10.1103/PhysRevLett.116.026402} {\bibfield  {journal} {\bibinfo  {journal}
  {Physical Review Letters}\ }\textbf {\bibinfo {volume} {116}},\ \bibinfo
  {eid} {026402} (\bibinfo {year} {2016})}\BibitemShut {NoStop}%
\bibitem [{\citenamefont {Motrunich}\ \emph {et~al.}(2001)\citenamefont
  {Motrunich}, \citenamefont {Damle},\ and\ \citenamefont
  {Huse}}]{motrunich_pwave}%
  \BibitemOpen
  \bibfield  {author} {\bibinfo {author} {\bibfnamefont {O.}~\bibnamefont
  {Motrunich}}, \bibinfo {author} {\bibfnamefont {K.}~\bibnamefont {Damle}}, \
  and\ \bibinfo {author} {\bibfnamefont {D.~A.}\ \bibnamefont {Huse}},\ }\href
  {\doibase 10.1103/PhysRevB.63.224204} {\bibfield  {journal} {\bibinfo
  {journal} {Phys. Rev. B}\ }\textbf {\bibinfo {volume} {63}},\ \bibinfo
  {pages} {224204} (\bibinfo {year} {2001})}\BibitemShut {NoStop}%
\bibitem [{\citenamefont {Cole}\ \emph {et~al.}(2016)\citenamefont {Cole},
  \citenamefont {Sau},\ and\ \citenamefont {Das~Sarma}}]{cole_disorder}%
  \BibitemOpen
  \bibfield  {author} {\bibinfo {author} {\bibfnamefont {W.~S.}\ \bibnamefont
  {Cole}}, \bibinfo {author} {\bibfnamefont {J.~D.}\ \bibnamefont {Sau}}, \
  and\ \bibinfo {author} {\bibfnamefont {S.}~\bibnamefont {Das~Sarma}},\ }\href
  {\doibase 10.1103/PhysRevB.94.140505} {\bibfield  {journal} {\bibinfo
  {journal} {Phys. Rev. B}\ }\textbf {\bibinfo {volume} {94}},\ \bibinfo
  {pages} {140505} (\bibinfo {year} {2016})}\BibitemShut {NoStop}%
\bibitem [{Note1()}]{Note1}%
  \BibitemOpen
  \bibinfo {note} {The $M$ dependence can be made arbitrarily small at fixed
  $mL$ by increasing $L$, in accordance with our expectation for a
  non-universal boundary condition contribution to the excitation
  energy.}\BibitemShut {Stop}%
\bibitem [{\citenamefont {G\"{u}l}\ \emph {et~al.}(2017)\citenamefont
  {G\"{u}l}, \citenamefont {Zhang}, \citenamefont {de~Vries}, \citenamefont
  {van Veen}, \citenamefont {Zuo}, \citenamefont {Mourik}, \citenamefont
  {Conesa-Boj}, \citenamefont {Nowak}, \citenamefont {van Woerkom},
  \citenamefont {Quintero-P\'{e}rez}, \citenamefont {Cassidy}, \citenamefont
  {Geresdi}, \citenamefont {Koelling}, \citenamefont {Car}, \citenamefont
  {Plissard}, \citenamefont {Bakkers},\ and\ \citenamefont
  {Kouwenhoven}}]{gul_hardgap}%
  \BibitemOpen
  \bibfield  {author} {\bibinfo {author} {\bibfnamefont {O.}~\bibnamefont
  {G\"{u}l}}, \bibinfo {author} {\bibfnamefont {H.}~\bibnamefont {Zhang}},
  \bibinfo {author} {\bibfnamefont {F.~K.}\ \bibnamefont {de~Vries}}, \bibinfo
  {author} {\bibfnamefont {J.}~\bibnamefont {van Veen}}, \bibinfo {author}
  {\bibfnamefont {K.}~\bibnamefont {Zuo}}, \bibinfo {author} {\bibfnamefont
  {V.}~\bibnamefont {Mourik}}, \bibinfo {author} {\bibfnamefont
  {S.}~\bibnamefont {Conesa-Boj}}, \bibinfo {author} {\bibfnamefont {M.~P.}\
  \bibnamefont {Nowak}}, \bibinfo {author} {\bibfnamefont {D.~J.}\ \bibnamefont
  {van Woerkom}}, \bibinfo {author} {\bibfnamefont {M.}~\bibnamefont
  {Quintero-P\'{e}rez}}, \bibinfo {author} {\bibfnamefont {M.~C.}\ \bibnamefont
  {Cassidy}}, \bibinfo {author} {\bibfnamefont {A.}~\bibnamefont {Geresdi}},
  \bibinfo {author} {\bibfnamefont {S.}~\bibnamefont {Koelling}}, \bibinfo
  {author} {\bibfnamefont {D.}~\bibnamefont {Car}}, \bibinfo {author}
  {\bibfnamefont {S.~R.}\ \bibnamefont {Plissard}}, \bibinfo {author}
  {\bibfnamefont {E.~P. A.~M.}\ \bibnamefont {Bakkers}}, \ and\ \bibinfo
  {author} {\bibfnamefont {L.~P.}\ \bibnamefont {Kouwenhoven}},\ }\href
  {\doibase 10.1021/acs.nanolett.7b00540} {\bibfield  {journal} {\bibinfo
  {journal} {Nano Letters}\ }\textbf {\bibinfo {volume} {17}},\ \bibinfo
  {pages} {2690} (\bibinfo {year} {2017})}\BibitemShut {NoStop}%
\bibitem [{\citenamefont {{Deng}}\ \emph {et~al.}(2016)\citenamefont {{Deng}},
  \citenamefont {{Vaitiek{\.e}nas}}, \citenamefont {{Hansen}}, \citenamefont
  {{Danon}}, \citenamefont {{Leijnse}}, \citenamefont {{Flensberg}},
  \citenamefont {{Nyg{\aa}rd}}, \citenamefont {{Krogstrup}},\ and\
  \citenamefont {{Marcus}}}]{deng_qd}%
  \BibitemOpen
  \bibfield  {author} {\bibinfo {author} {\bibfnamefont {M.~T.}\ \bibnamefont
  {{Deng}}}, \bibinfo {author} {\bibfnamefont {S.}~\bibnamefont
  {{Vaitiek{\.e}nas}}}, \bibinfo {author} {\bibfnamefont {E.~B.}\ \bibnamefont
  {{Hansen}}}, \bibinfo {author} {\bibfnamefont {J.}~\bibnamefont {{Danon}}},
  \bibinfo {author} {\bibfnamefont {M.}~\bibnamefont {{Leijnse}}}, \bibinfo
  {author} {\bibfnamefont {K.}~\bibnamefont {{Flensberg}}}, \bibinfo {author}
  {\bibfnamefont {J.}~\bibnamefont {{Nyg{\aa}rd}}}, \bibinfo {author}
  {\bibfnamefont {P.}~\bibnamefont {{Krogstrup}}}, \ and\ \bibinfo {author}
  {\bibfnamefont {C.~M.}\ \bibnamefont {{Marcus}}},\ }\href {\doibase
  10.1126/science.aaf3961} {\bibfield  {journal} {\bibinfo  {journal}
  {Science}\ }\textbf {\bibinfo {volume} {354}},\ \bibinfo {pages} {1557}
  (\bibinfo {year} {2016})}\BibitemShut {NoStop}%
\bibitem [{\citenamefont {{Chen}}\ \emph {et~al.}(2016)\citenamefont {{Chen}},
  \citenamefont {{Yu}}, \citenamefont {{Stenger}}, \citenamefont {{Hocevar}},
  \citenamefont {{Car}}, \citenamefont {{Plissard}}, \citenamefont {{Bakkers}},
  \citenamefont {{Stanescu}},\ and\ \citenamefont
  {{Frolov}}}]{frolov_phasediagram}%
  \BibitemOpen
  \bibfield  {author} {\bibinfo {author} {\bibfnamefont {J.}~\bibnamefont
  {{Chen}}}, \bibinfo {author} {\bibfnamefont {P.}~\bibnamefont {{Yu}}},
  \bibinfo {author} {\bibfnamefont {J.}~\bibnamefont {{Stenger}}}, \bibinfo
  {author} {\bibfnamefont {M.}~\bibnamefont {{Hocevar}}}, \bibinfo {author}
  {\bibfnamefont {D.}~\bibnamefont {{Car}}}, \bibinfo {author} {\bibfnamefont
  {S.~R.}\ \bibnamefont {{Plissard}}}, \bibinfo {author} {\bibfnamefont
  {E.~P.~A.~M.}\ \bibnamefont {{Bakkers}}}, \bibinfo {author} {\bibfnamefont
  {T.~D.}\ \bibnamefont {{Stanescu}}}, \ and\ \bibinfo {author} {\bibfnamefont
  {S.~M.}\ \bibnamefont {{Frolov}}},\ }\href@noop {} {\bibfield  {journal}
  {\bibinfo  {journal} {ArXiv e-prints}\ } (\bibinfo {year} {2016})},\ \Eprint
  {http://arxiv.org/abs/1610.04555} {arXiv:1610.04555 [cond-mat.mes-hall]}
  \BibitemShut {NoStop}%
\bibitem [{\citenamefont {{Das Sarma}}\ \emph {et~al.}(2012)\citenamefont {{Das
  Sarma}}, \citenamefont {{Sau}},\ and\ \citenamefont
  {{Stanescu}}}]{smokinggun}%
  \BibitemOpen
  \bibfield  {author} {\bibinfo {author} {\bibfnamefont {S.}~\bibnamefont {{Das
  Sarma}}}, \bibinfo {author} {\bibfnamefont {J.~D.}\ \bibnamefont {{Sau}}}, \
  and\ \bibinfo {author} {\bibfnamefont {T.~D.}\ \bibnamefont {{Stanescu}}},\
  }\href {\doibase 10.1103/PhysRevB.86.220506} {\bibfield  {journal} {\bibinfo
  {journal} {\prb}\ }\textbf {\bibinfo {volume} {86}},\ \bibinfo {eid} {220506}
  (\bibinfo {year} {2012})}\BibitemShut {NoStop}%
\bibitem [{\citenamefont {{Rainis}}\ \emph {et~al.}(2013)\citenamefont
  {{Rainis}}, \citenamefont {{Trifunovic}}, \citenamefont {{Klinovaja}},\ and\
  \citenamefont {{Loss}}}]{diego_realistic}%
  \BibitemOpen
  \bibfield  {author} {\bibinfo {author} {\bibfnamefont {D.}~\bibnamefont
  {{Rainis}}}, \bibinfo {author} {\bibfnamefont {L.}~\bibnamefont
  {{Trifunovic}}}, \bibinfo {author} {\bibfnamefont {J.}~\bibnamefont
  {{Klinovaja}}}, \ and\ \bibinfo {author} {\bibfnamefont {D.}~\bibnamefont
  {{Loss}}},\ }\href {\doibase 10.1103/PhysRevB.87.024515} {\bibfield
  {journal} {\bibinfo  {journal} {\prb}\ }\textbf {\bibinfo {volume} {87}},\
  \bibinfo {eid} {024515} (\bibinfo {year} {2013})}\BibitemShut {NoStop}%
\bibitem [{\citenamefont {Zyuzin}\ \emph {et~al.}(2013)\citenamefont {Zyuzin},
  \citenamefont {Rainis}, \citenamefont {Klinovaja},\ and\ \citenamefont
  {Loss}}]{zyuzin}%
  \BibitemOpen
  \bibfield  {author} {\bibinfo {author} {\bibfnamefont {A.~A.}\ \bibnamefont
  {Zyuzin}}, \bibinfo {author} {\bibfnamefont {D.}~\bibnamefont {Rainis}},
  \bibinfo {author} {\bibfnamefont {J.}~\bibnamefont {Klinovaja}}, \ and\
  \bibinfo {author} {\bibfnamefont {D.}~\bibnamefont {Loss}},\ }\href {\doibase
  10.1103/PhysRevLett.111.056802} {\bibfield  {journal} {\bibinfo  {journal}
  {Phys. Rev. Lett.}\ }\textbf {\bibinfo {volume} {111}},\ \bibinfo {pages}
  {056802} (\bibinfo {year} {2013})}\BibitemShut {NoStop}%
\bibitem [{\citenamefont {Akhmerov}\ \emph {et~al.}(2011)\citenamefont
  {Akhmerov}, \citenamefont {Dahlhaus}, \citenamefont {Hassler}, \citenamefont
  {Wimmer},\ and\ \citenamefont {Beenakker}}]{akhmerov_QCP}%
  \BibitemOpen
  \bibfield  {author} {\bibinfo {author} {\bibfnamefont {A.~R.}\ \bibnamefont
  {Akhmerov}}, \bibinfo {author} {\bibfnamefont {J.~P.}\ \bibnamefont
  {Dahlhaus}}, \bibinfo {author} {\bibfnamefont {F.}~\bibnamefont {Hassler}},
  \bibinfo {author} {\bibfnamefont {M.}~\bibnamefont {Wimmer}}, \ and\ \bibinfo
  {author} {\bibfnamefont {C.~W.~J.}\ \bibnamefont {Beenakker}},\ }\href
  {\doibase 10.1103/PhysRevLett.106.057001} {\bibfield  {journal} {\bibinfo
  {journal} {Phys. Rev. Lett.}\ }\textbf {\bibinfo {volume} {106}},\ \bibinfo
  {pages} {057001} (\bibinfo {year} {2011})}\BibitemShut {NoStop}%
\bibitem [{\citenamefont {Fregoso}\ \emph {et~al.}(2013)\citenamefont
  {Fregoso}, \citenamefont {Lobos},\ and\ \citenamefont
  {Das~Sarma}}]{fregoso_QCP}%
  \BibitemOpen
  \bibfield  {author} {\bibinfo {author} {\bibfnamefont {B.~M.}\ \bibnamefont
  {Fregoso}}, \bibinfo {author} {\bibfnamefont {A.~M.}\ \bibnamefont {Lobos}},
  \ and\ \bibinfo {author} {\bibfnamefont {S.}~\bibnamefont {Das~Sarma}},\
  }\href {\doibase 10.1103/PhysRevB.88.180507} {\bibfield  {journal} {\bibinfo
  {journal} {Phys. Rev. B}\ }\textbf {\bibinfo {volume} {88}},\ \bibinfo
  {pages} {180507} (\bibinfo {year} {2013})}\BibitemShut {NoStop}%
\bibitem [{\citenamefont {Tewari}\ \emph {et~al.}(2012)\citenamefont {Tewari},
  \citenamefont {Sau}, \citenamefont {Scarola}, \citenamefont {Zhang},\ and\
  \citenamefont {Das~Sarma}}]{tewari_QCP}%
  \BibitemOpen
  \bibfield  {author} {\bibinfo {author} {\bibfnamefont {S.}~\bibnamefont
  {Tewari}}, \bibinfo {author} {\bibfnamefont {J.~D.}\ \bibnamefont {Sau}},
  \bibinfo {author} {\bibfnamefont {V.~W.}\ \bibnamefont {Scarola}}, \bibinfo
  {author} {\bibfnamefont {C.}~\bibnamefont {Zhang}}, \ and\ \bibinfo {author}
  {\bibfnamefont {S.}~\bibnamefont {Das~Sarma}},\ }\href {\doibase
  10.1103/PhysRevB.85.155302} {\bibfield  {journal} {\bibinfo  {journal} {Phys.
  Rev. B}\ }\textbf {\bibinfo {volume} {85}},\ \bibinfo {pages} {155302}
  (\bibinfo {year} {2012})}\BibitemShut {NoStop}%
\end{thebibliography}



%


\end{document}